\documentclass[aps,prc,preprint,groupedaddress,showpacs]{revtex4}

\usepackage{graphicx}
\begin{document}


\title{Triple-Pomeron Matrix Model for Dispersive Corrections to\\
Nucleon-Nucleus Total Cross Section}


\author{David R. Harrington}
\email[]{drharr@physics.rutgers.edu}
\affiliation{Deparment of Physics and Astronomy, Rutgers
University \\  136 Frelinghuysen Road \\  Piscataway, NJ
08854-8019, USA }


\date{\today}

\begin{abstract}
Dispersive corrections to the total cross section for high-energy
scattering from a heavy nucleus are calculated using a matrix
model, based on the triple-Pomeron behavior of diffractive
scattering from a single nucleon, for the cross section operator
connecting different states of the projectile nucleon .
Energy-dependent effects due to the decrease in longitudinal
momentum transfers and the opening of more channels with
increasing energy are included.  The three leading terms in an
expansion in the number of inelastic transitions are evaluated
and compared to exact results for the model in the uniform nuclear
density approximation for the the scattering of nucleons from
$Pb^{208}$ for laboratory momenta ranging
 from 50 to 200 GeV/c   . 
\end{abstract}


\maketitle

\section{INTRODUCTION}
One consequence of the composite nature of nucleons is a decrease
in nucleon-nucleus total cross sections due to transitions between
different internal states of the  projectile nucleon. This
decrease can be calculated \cite{bla,har} using an operator to
represent a generalized nucleon-nucleon total cross section, with
the matrix elements representing the probability amplitudes for
forward scattering transitions between different states of the
nucleon.

    The general problem of the propagation of composite particles through
a nucleus has a long history
\cite{GandW,drell,ross,KandM,vanH,huf} .   If the longitudinal
momentum transfer due to the different masses of the different
states of the composite system cannot be ignored  the calculation
of the dispersive corrections becomes quite complicated.  Because
of the difficulty of this calculation, and uncertainties in the
nature of the cross section operator, only the leading term in an
expansion in the number of inelastic transitions has been
evaluated \cite{jen}. (A closely related effect in quasi-elastic
electron scattering can complicate the analysis of color
transparency effects in these reactions \cite{kop}.) In this
paper we  test the accuracy of this approximation, and the
convergence of the expansion, using a simple finite matrix model
for the cross section operator and taking the uniform nuclear
density limit for which the exact result, including all orders in
transitions, can also be calculated. This matrix model is
consistent with the known triple-Pomeron behavior of high-energy,
high-mass single diffraction dissociation from a single nucleon,
but otherwise has a high degree of arbitrariness:  it is used only
for lack of a reliable microscopic model for the internal degrees
of freedom of highly excited nucleons.    It has some features in
common with a "simplified example" used by Van Hove \cite{vanH}
for the limit of zero longitudnal momentum transfer.

    Section II reviews the  transition expansion for the
dispersive corrections to the nucleon-nucleus total cross section,
including the effects of the longitudinal momentum transfers due
to the different masses of the the different excited states of the
nucleon.   In Sect. III it is shown that the terms in this
expansion simplify considerably in the uniform nuclear density
limit, with  each term in the expansion  represented as a sum
over products of transition amplitudes weighted by a function of
the differences among cross sections and differences in the
masses of the different nucleonic states. In the uniform density
limit it is also possible to write  the exact result, including
all orders in the number of transitions, in terms of the
exponential of a single position-independent operator, as shown
in Sect. IV. The representation of the transition operator  by a
finite matrix, with a dimension which increases with energy, is
discussed in Sect. V.   This matrix is chosen to have a form
consistent with the triple-Pomeron behavior of the
nucleon-nucleon single diffraction dissociation, but is otherwise
highly arbitrary. Using this matrix the
 formulas  are evaluated and the results presented in
Sect. VI.  The results are summarized and discussed in the
concluding Sect. VII.

\section{Expansion in Inelastic Transitions}

    It has long been known that a version of the eikonal
approximation holds for an infinite but restricted class of
Feynman diagrams which includes inelastic transitions among
different states of the projectile \cite{harr}.  The result is
equivalent to the eikonal approximation in coupled-channel
potential theory\cite{harr68,BandM,FandH}, and leads to an
expression for projectile-nucleus total cross sections which can
be written as
\begin{equation}
\sigma(A)_{Total} = 2  Re \int d^2b \langle A | \langle 1 |
\hat{\Gamma }({\bf b},\{ {\bf r}_{\alpha} \} ) | 1 \rangle |A
\rangle ,
\end{equation}
where $ | A \rangle $ is the ground state of the nucleus and  $ |
1 \rangle$ is the lowest mass eigenstate of the projectile
system.  The profile function ${\hat \Gamma} $ is an operator in
the internal space of the projectile:
\begin{equation}
     {\hat \Gamma}({\bf b},\{ {\bf r}_{\alpha} \}) =1-{\cal
Z}exp[-i \sum_{\alpha} \int dz {\hat u}_{\alpha}({\bf r}) ] ,
\end{equation}
where $\cal Z$ indicates a z-ordered product and
\begin{equation}
       {\hat u}_{\alpha}({\bf r}) = (m/p_1 ) e^{-i\hat{p} z}{\hat v}({\bf r-r_{\alpha}}) e^{i\hat{p}
       z},
\end{equation}
with ${\hat v}({\bf r-r_{\alpha} }) $  the effective potential
operator produced by a static target nucleon at $ {\bf r_{\alpha}}
$. Here  $ \hat{p}$ is the longitudinal momentum operator,
diagonal in the mass eigenstates of the projectile, with
\begin{equation}
    p_{j} \approx p_{1} - (m_{j}^2 -m_{1}^2)/(2p_1),
\end{equation}
where $p_1$ is the initial momentum of the projectile in the
laboratory system where the nucleus is at rest and $m_j$ is the
mass of the $jth$ excited state of the projectile.

    Assuming the different ${\hat v_{\alpha}}$ do not overlap,
ignoring nuclear correlations, and assuming a large nucleon number
 $A$, the expression for ${\hat \Gamma}$ simplifies to
\begin{equation}
    {\hat \Gamma}(b) = \langle A |{\hat \Gamma}(b);\{{\bf
r_{\alpha}} \}|A \rangle \\
            \approx 1 - {\cal Z} exp[-(A/2) \int dz_1 \rho
(b,z_1 ) {\hat \sigma } (z_1) ],
\end{equation}
where $\rho $ is the nuclear density, normalized to one, and
\begin{equation}
    {\hat \sigma}(z_1) \equiv exp(-i{\hat p}z_1){\hat \sigma} exp(i{\hat p}z_1),
\end{equation}
with  ${\hat \sigma}$  the cross section operator \cite{har} for
scattering of the projectile from a single target nucleon.
Assuming that the corresponding elastic and diffractive scattering
amplitudes are purely imaginary, the total projectile-nucleon
cross section is
\begin{equation}
    \sigma _{T} = \langle 1 |{\hat \sigma} |1 \rangle ,
\end{equation}
while the  cross section for single diffraction dissociation of
the projectile interacting with a single nucleon, summed over all
diffractively excited states of the projectile, at momentum
transfer squared $t = 0$, is
\begin{equation}
 d\sigma _{diff}/dt = [\langle 1 |{\hat \sigma}^{2} |1 \rangle -  \langle 1
|{\hat \sigma} |1 \rangle ^{2} ]/(16\pi ).
\end{equation}
If
\begin{equation}
    \Gamma (b) \equiv \langle 1 | {\hat \Gamma }(b) | 1 \rangle  ,
\end{equation}
then
\begin{equation}
\sigma(A)_{Total} = 2  Re \int d^2b \, \Gamma (b).
\end{equation}

    Since the dispersive corrections  are
small compared to the total cross section it is useful to separate
$\Gamma (b) $ into two parts:
\begin{equation}
 \Gamma (b)   = \Gamma _{G}(b)
-\Gamma _{D}(b) .
\end{equation}
Defining the dimensionless absorption parameter
\begin{equation}
    t(b) = (A/2) \sigma _{T} \int dz \rho (b,z) ,
\end{equation}
 the main Glauber contribution, which does not include
dispersive corrections, is simply
\begin{equation}
    \Gamma _{G} (b) = 1 - exp[-t(b)],
\end{equation}
while the diffractive correction is given by
\begin{equation}
    \Gamma _{D}(b) = \langle 1 |{\cal Z}exp[-(A/2) \int dz \rho
(b,z) {\hat \sigma }(z)] |1 \rangle \ - exp[-t(b)] .
\end{equation}

    Previous calculations taking into account longitudnal momentum
 transfers \cite{jen} \, of $\Gamma _{D}$ have included only the
leading second order term in an expansion in the number of
inelastic transition.  While this is almost certainly accurate for
light nuclei, it is not clear whether or not it is adequate for
heavy nuclei. Below we develop expressions for the general terms
in the expansion and evaluate them in a simple but possibly not
completely unrealistic model.

    We begin by separating the cross section operator ${\hat
\sigma}$ into its diagonal and off-diagonal (transition) parts:
\begin{equation}
    {\hat \sigma} = {\hat \sigma}_{d} +  {\hat \sigma}_{t},
\end{equation}
where
\begin{equation}
    \langle i |{\hat \sigma}_{d} |j\rangle = \delta
_{ij}\langle j|{\hat  \sigma} |j\rangle ,
\end{equation}
so that ${\hat \sigma}_{t} $ has only off-diagonal matrix
elements.  We now expand $\Gamma _D $ in powers of the transition
operator:
\begin{equation}
    \Gamma _{D}(b) = \sum_{n=2}^{\infty } \Gamma _{D}^{(n)}(b) .
\end{equation}
Here  the leading  $n=2$ term contains contributions from
processes in which the nucleon makes two transitions: one from the
ground state to a higher-mass state, then another back to the
ground state. The next ($n=3$ ) term contains the contributions
from processes with three transitions, with the two intermediate
projectile states being neither the ground state nor equal to each
other.

    Using a derivation analogous to that for time-dependent
perturbation theory \cite{shankar} one can show that the $\Gamma
_{D}^{(n)}(b)$ are given by the z-ordered integrals
\begin{eqnarray}
    \Gamma _{D}^{(n)}(b) &=& (-A/2)^{n} e^{-t}
\int_{-\infty}^{\infty}dz_{n}\rho (b,z_{n}) \ldots
\int_{-\infty}^{z_2} dz_1 \rho (b,z_1) \nonumber \\
     &  & \langle
1|{\hat \sigma}_{td}(z_{n}) \ldots {\hat \sigma}_{td}(z_{2}){\hat
\sigma}_{td}(z_{1}) |1 \rangle ,
\end{eqnarray}
with
\begin{equation}
{\hat\sigma}_{td}(z) = {\hat U}(z)^{-1}\, {\hat \sigma}_{t}\,{\hat
U}(z),
\end{equation}
where
\begin{equation}
    {\hat U}(z) = exp[ - (A/2) \int_{-\infty}^{z}dz_1 \rho(b,z_1 )
{\hat \sigma}_{d} + i{\hat p}\, z ]
\end{equation}
 is diagonal in mass eigenstates. These expressions can be simplified if we replace $z$ by the
dimensionless variable
\begin{equation}
    u(b,z) \equiv (A/2)\sigma _T  \int_{-\infty}^{z}dz_1
\rho(b,z_1 ) /t(b),
\end{equation}
so that $u(b,-\infty ) = 0$ and $u(b, \infty ) = 1$.  If we also
define the dimensionless cross section operators
\begin{equation}
    {\hat x} \equiv {\hat \sigma } / \sigma _T,
\end{equation}
\begin{equation}
    {\hat x}_{d} \equiv {\hat \sigma }_{d} / \sigma _T,
\end{equation}
 \begin{equation}
    {\hat x}_{t} \equiv {\hat \sigma }_{t} / \sigma _T ,
\end{equation}
and, for future use,
\begin{equation}
    {\tilde x} \equiv {\hat x} - {\hat 1}.
\end{equation}
 We can  then write
\begin{eqnarray}
    \Gamma_{D}^{(n)}(b) &=& (-t(b))^{n} exp[-t(b)] \int_{0}^{1}du_{n}
\ldots  \int_{0}^{u_3}du_{2}  \int_{0}^{u_2}du_{1} \nonumber \\
      &  &  \langle 1| {\hat x}_{td}(u_{n})  \ldots {\hat x}_{td}(u_{1}) |1
\rangle ,
\end{eqnarray}
where
\begin{equation}
    {\hat x}_{td}(u) = {\hat U}(z(b,u))^{-1}\, {\hat x}_{t}\, {\hat U}(z(b,u)) ,
\end{equation}
with $z(b,u)$  the inverse of $u(b,z)$ for fixed $b$.

    Inserting complete sets of mass eigenstates between the ${\hat
x}_{td}(u)$ operators in Eqn. (20) gives
\begin{eqnarray}
        \Gamma_{D}^{(n)}(b) & = & [(-t(b))^{n}/n ! ] exp[-t(b)]
\sum_{j_1,\ldots ,j_{n-1}} \langle 1| {\hat x}_{t} | j_{n-1}
\rangle \ldots \langle j_2| {\hat x}_{t} | j_1 \rangle  \langle
j_1| {\hat x}_{t} | 1 \rangle \, \nonumber \\
 &  &  f^{(n)} (b;j_1, j_2, \ldots ,j_{n-1}) ,
\end{eqnarray}
where
\begin{eqnarray}
    f^{(n)} (b; j_1, j_2, \ldots ,j_{n-1}) &  \equiv & n !  \int_{0}^{1}du_{n}
\ldots  \int_{0}^{u_3}du_{2}  \int_{0}^{u_2}du_{1}   \nonumber
\\ & & exp[t(b) u_{n}
(x_1-x_{j_{n-1}})-iz_n (p_1 -p_{j_{n-1}})]   \nonumber \\
  & & \ldots  exp[t(b) u_{2}
(x_{j_2}-x_{j_{1}})-iz_2 (p_{j_2}-p_{j_{1}})] \nonumber \\
    &  & exp[t(b) u_{1} (x_{j_1}-x_{1})-iz_1 (p_{j_1}-p_{1})] ,
\end{eqnarray}
with $z_j=z(b,u_j) $ and $x_j = <j|{\hat x}|j>$.  Since ${\hat
x}_t$ has only off-diagonal matrix elements, terms with equal
successive $ j_i \, 's$ do not contribute to the sum in Eqn 22.

    Clearly the functions $f^{(n)}$ defined above depend only on
the differences between successive $x_j 's$ and $p_j 's$, and
thus are unchanged if  these variables are replaced by
\begin{equation}
    \tilde{x}_j \equiv x_j -1
\end{equation}
and
\begin{equation}
    \tilde{p}_j \equiv p_j -p_1  .
\end{equation}
Then $f^{(n)} $ can be written
\begin{eqnarray}
    f^{(n)} (b; j_1, j_2, \ldots ,j_{n-1}) & =   &   n !  \int_{0}^{1}du_{n}
\ldots  \int_{0}^{u_3}du_{2}  \int_{0}^{u_2}du_{1}  \nonumber  \\
        &  &    exp[-t
\tilde{x}_{j_{n-1}} (u_n - u_{n-1}) + i \tilde{p}_{j_{n-1}} (z_n -
z_{n-1})]  \nonumber  \\
 & & \ldots   exp[-t
\tilde{x}_{j_1} (u_2 - u_{1}) + i \tilde{p}_{j_{2}} (z_{2} -
z_{1})] ,
\end{eqnarray}
where, because of the u (and z) ordering, the differences between
$u's$ and $z's$ in the parenthesis are never negative, and the
real parts of the $ f^{(n)}\, 's $, needed to calculate the total
cross sections, are always less than their limits as the
$\tilde{p} _{j} \, 's$ approach zero.  Each exponential in this
expression acts as a propagator for the projectile from the
location of one transition to that of the next and includes
absorptive and phase-changing parts depending upon the state of
the projectile at this stage of its journey through the nucleus.

\section{Uniform Density Limit}

    The  evaluation of the expressions for the $f^{(n)}\, 's$ is complicated by the
fact that $z$ and $u$ are in general not simple functions of one
another: the relation between them is determined by the shape of
the nuclear density function $\rho $ and depends upon the impact
parameter. For heavy nuclei $\rho $ is well-approximated by the
simple Woods-Saxon form, but the relation between $z$ and $u$ is
still not simple.  For the heaviest nuclei the surface thickness
is much less than the nuclear radius so that it may not introduce
excessive errors to replace $\rho $ by its uniform density limit,
especially at small momentum transfers and , in particular, for
evaluating the dispersive contribution to the total cross
section.  In this limit
\begin{equation}
    \rho (r) = \rho _{0} \Theta (R-r) ,
\end{equation}
giving
\begin{equation}
    t(b)= A\sigma _T \rho _0 \sqrt{R^2-b^2}\, \Theta (R-b),
\end{equation}
where
\begin{equation}
    \rho _0 =1/(4\pi R^3 /3)
\end{equation}
and  $R \approx r_0 A^{1/3} $ , with $r_0 \approx 1.14 fm $ .
(With this expression for $t(b)$ there is an analytic expression
\cite{drell} for the main Glauber contribution  to $\sigma
(A)_{Total} $.)
 In this limit $z$ and $u$ are linearly related:
\begin{equation}
    z = \sqrt{R^2 - b^2 } (2u-1)
\end{equation}
for $ 0 \leq u \leq 1$.  The arguments of the exponentials in Eqn.
26  then simplify considerably:
\begin{equation}
-\tilde{x}_{j_n} t(b)(u_{n+1}-u_n) + i \tilde{p}_{j_n} (z_{n+1}
-z_n) = -y_{j_n} \sqrt{R^2 - b^2} (u_{n+1} -u_n) ,
\end{equation}
where the complex number
\begin{equation}
    y_j \equiv \tilde{x}_j A \sigma _T \rho _0  -2i\tilde{p}_j
\end{equation}
is independent of $b$ and the $u's$ .  In the uniform density
limit, then, the effects of longitudinal momentum transfers are
taken into account simply by adding an imaginary part to each
diagonal matrix element $\tilde{x}_j$, and modifying the
calculations of the functions $f^{(n)}$ accordingly.

\section{Exact Result}

    In the uniform density limit an exact expression for  the
dispersive correction correction  to the total cross section can
be found in terms of the exponential of a z-independent operator.
The simplest derivation of this result starts with Eqn. 5 and
removes the z-dependence of $\hat{\sigma} (z)$ by adding a term
proportional to $\hat{p}$ to the operator in the exponent. For a
given impact parameter a single z-independent matrix is involved
so that the z-ordering in Eqn. (2) can be ignored, giving
\begin{equation}
    \Gamma _D (b) = \langle 1 | exp(-\hat{M}(b) |1\rangle -exp(-t(b)) ,
\end{equation}
with the z-independent operator
\begin{equation}
    \hat{M} (b) = [ A \rho _0 \hat{\sigma} - 2i(\hat{p}-p_1) ]
    \sqrt{R^2 - b^2}.
\end{equation}
This can also be written as
\begin{equation}
    \Gamma _D (b) = exp(-t(b)) [\langle 1 | exp(-\tilde{M}(b)) |1\rangle
    - 1] ,
\end{equation}
where
\begin{equation}
    \tilde{M}(b) = \hat{M} (b) - t(b)\hat{1} =[ A \rho _0 \sigma _T \tilde{x} -
    2i\tilde{p}]
    \sqrt{R^2 - b^2}.
\end{equation}
Either of the expressions (33) or (35)can be evaluated by
expanding the exponential of the z-independent operator in a
power series , with the second converging somewhat more rapidly.
(Using the matrix model below one has to include of the order of
50 terms in the expansion, and there is considerable
cancellation, so the individual terms must be calculated to high
accuracy.)

    Another approach for evaluating $\Gamma _D (b)$ depends upon the fact that, since the nucleon always
enters the reaction in its ground state, the full operator
$exp(-\tilde{M})$ is not needed:  it is sufficient to work with
the reaction-modified state
\begin{equation}
    |V\rangle \equiv exp(-\tilde{M}) |1\rangle ,
\end{equation}
with
\begin{equation}
    \Gamma _D (b) = exp(-t(b)) [\langle 1 |   V \rangle
    - 1] .
\end{equation}
Expanding the exponential
\begin{equation}
    |V\rangle =  \sum_{n=0}^{\infty} |Vn\rangle ,
\end{equation}
where
\begin{equation}
    |Vn\rangle \equiv (1/n!)(-\tilde{M})^n |1\rangle = -(1/n) \tilde{M}|V(n-1)\rangle
\end{equation}
can be calculated recursively starting with $|V0\rangle =
|1\rangle $.

\section{Matrix Model}

    The expressions for $\Gamma_{D}^{(n)}$ above involve the
matrix elements of the dimensionless cross section operator
${\hat x}$ and the longitudinal momentum operator ${\hat p}$ . In
this section a model for these operators is developed which,
although highly arbitrary, is consistent with the experimental
high-energy behavior of diffraction dissociation, which is in turn
approximately consistent with that expected from  the leading
triple-Pomeron behavior.  Ignoring contributions from secondary
Regge poles, and taking the Pomeron intercept $\alpha (0) = 1$,
this leads to the simple behavior at momentum transfer squared
$t=0$ \cite{goul.rep, BandP}
\begin{equation}
    d\sigma /(dt dM^2) = \sigma _T \, ^{3/2} g_{PPP} /(16 \pi M^2 ),
\end{equation}
where $\sigma _T$ is the nucleon-nucleon total cross section and
$g_{PPP} \approx 0.364 mb^{1/2} $ \cite{cool} \, is the
triple-Pomeron vertex, while the mass-square of the diffractively
excited nucleon runs from $M_{min}^{2} \approx 1.5 ( GeV/c^2)^2 $
to $M_{max}^2 \approx m_1^2 +2 p_1 m_{\pi}   $ , the later
condition following from the requirement that the longitudinal
momentum transfer be less than $m_{\pi} $, the inverse of the
range of the strong force.  (At very high energies the effective
Pomeron intercept is greater than one and the above must be
modified.  Details can be found in \cite{tan,goul99}.)

    The continuous range of $m^2$ between $M_{min}^2 $ and $M_{max}^2$ can be approximately replaced by a
finite number of ``bins'' of width $\Delta m_j^2 $ centered at
$m_j^2$. The operator ${\hat x}$ can then be represented by a
finite matrix with elements $ \langle j | {\hat x} | i \rangle $
constrained by
\begin{equation}
    d\sigma _j /dt \approx \sigma _T^2 \langle j|{\hat x}|1\rangle ^2
/(16\pi)  \approx  [d\sigma /(dt dM^2)]_{m_j ^2} \Delta m_j ^2 ,
\end{equation}
or
\begin{equation}
    \langle j|{\hat x} |1\rangle ^2 \approx g_{PPP} \Delta m_j^2
/(\sigma _T^{1/2} m_j ^2)
\end{equation}
 To complete the model
one must also have a prescription for the sizes of the mass
bins.  For simplicity
 equal spacing in $m^2$ is used below:
\begin{equation}
    m_j^2 = m_1^2 + m_0^2 (j-1),
\end{equation}
with $m_0$ a parameter determined by the spacing of the low energy
diffractively produced resonances.  Eqn. (35) then takes the
simple form
\begin{equation}
    \langle j|{\hat x} |1\rangle ^2 \approx (g_{PPP}/\sigma
    _T^{1/2})/[(m_1/m_0)^2 +j-1].
\end{equation}
With  the expression  above for $M_{max}^2$, the dimension $N$ of
the matrix is  given by
\begin{equation}
    N \approx 2 p_1 m_{\pi}/m_0^2
\end{equation}
 which  increases linearly with $p_1$,  the
 momentum of the incident proton in the rest frame of the
target nucleus.  This expression for $N$ is clearly only a rough
estimate, but fortunately excited states with $j$ near $N$ do not
contribute much to the dispersive correction compared to lower
states:  changing $N$ slightly does not affect the results below
appreciably.
    Unfortunately diffraction dissociation constrains only one row (and column, from
the assumed symmetry) of the matrix $ \langle j|{\hat x}
|i\rangle $ .  For small $ m_i^2 $ and large $ m_j^2 $ one can
argue that the triple-Pomeron behavior should still be valid and
the $ \Delta m_j^2 / m_j^2 $ dependence should still hold.  For
simplicity, here it is assumed that {\em every} off-diagonal
element of the matrix is given by
\begin{equation}
     \langle j|{\hat x} |i\rangle = \sqrt{(g_{PPP}/[\sigma
_T^{1/2}(a^2 + |j-i|)]},
\end{equation}
where $a=m_1/m_0$ .  This expression
 is consistent with both experiment and the triple-Pomeron
behavior for $i=1$ and large $j$, but is only a guess elsewhere,
especially when $i$ and $j$ are comparable in size. Furthermore,
although the matrix is in general
 complex, it will below be assumed real.  This  is done
 mainly because the phases of the matrix elements are
 unknown (except for that of $x_{11}$) and is consistent with  the
 fact that the real part of the forward
 proton-proton scattering amplitude at high energies is known to be small.

    Finally, we need an expression for the diagonal elements $x_j=
\langle j|{\hat x} |j\rangle $ :
\begin{equation}
    x_j = 1 +  d (j-1) ,
\end{equation}
which allows the cross section for nucleons to scatter from
excited nucleons to increase with the degree of excitation. This
means that highly excited states will be absorbed more strongly
than lower states while propagating between transitions. (If $d=0$
all diagonal elements of $x$ are unity and and the expressions
above simplify considerably.) It would probably be more
reasonable for $x_j$ to approach a limiting value  as $j$
increases, but this would introduce still more parameters and
assumptions into the model.

    With these assumptions  both $x_j$ and $m_j^2$ are linear in
$j$, and so are
 the complex numbers $y_j$ defined in Eqn. 34:
\begin{equation}
    y_j = (j-1) y_2,
\end{equation}
where
\begin{equation}
    y_2 = A\sigma_T \rho_0 d + im_0^2/p_1  .
\end{equation}
For small $j's$ the influence of the longitudinal momentum
transfer, given by the imaginary part of $y_2$, decreases as
$1/p_1$, but the dimension  of the matrix increases as $p_1$ so
that for the heaviest excited nucleon included we have
\begin{equation}
    y_N = 2A\sigma_T \rho_0 d m_\pi p_1/m_0^2 +i2m_\pi ,
\end{equation}
with an imaginary part which is independent of $p_1$ and a real
part which increases linearly with momentum.

\section{Results}

    The formulas above have been evaluated for scattering of
nucleons from   $^{208}Pb$ for  incident laboratory momenta from
50 to 200 GeV/c,  and for $n$, the number of inelastic
diffractive transitions, ranging from 2 to 4. The corresponding
reductions in the total cross sections can be written as
\begin{equation}
    \sigma _D  ^{(n)} = \int db \, 4\pi b \, Re \Gamma _D ^{(n)} (b).
\end{equation}
Using $g_{PPP} = 0.363 mb^{1/2}$ and $\sigma_T= 38.5 mb$ gives the
coefficient in Eqn. 39 $(g_{PPP}/\sigma_T^{1/2})^{1/2}= 0.24$. We
also take $m_0^2=1.5GeV^2$, $a^2 = 0.5$ and $d= 0.1$, although
any values of the same order of magnitude would be just as
reasonable.
\begin{figure}
    \includegraphics{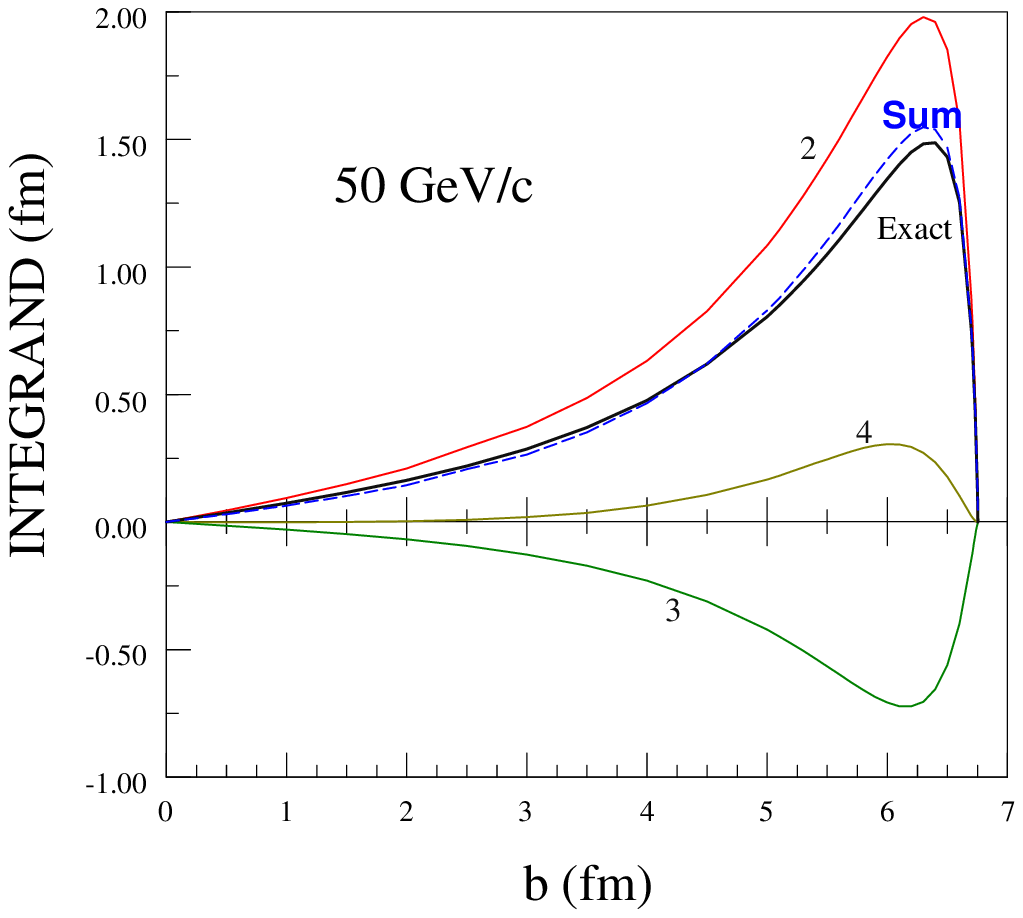}
    \caption{ Integrands for diffractive reductions in
    nucleon-$Pb^{208}$total cross sections as a function of impact parameter $b$
     at 50 GeV in 2nd through 4th order in the number of inelastic transitions.  The dashed  curve
    shows the sum of these corrections, while the heavy solid curve shows the exact complete correction.}
\end{figure}

\begin{figure}
    \includegraphics{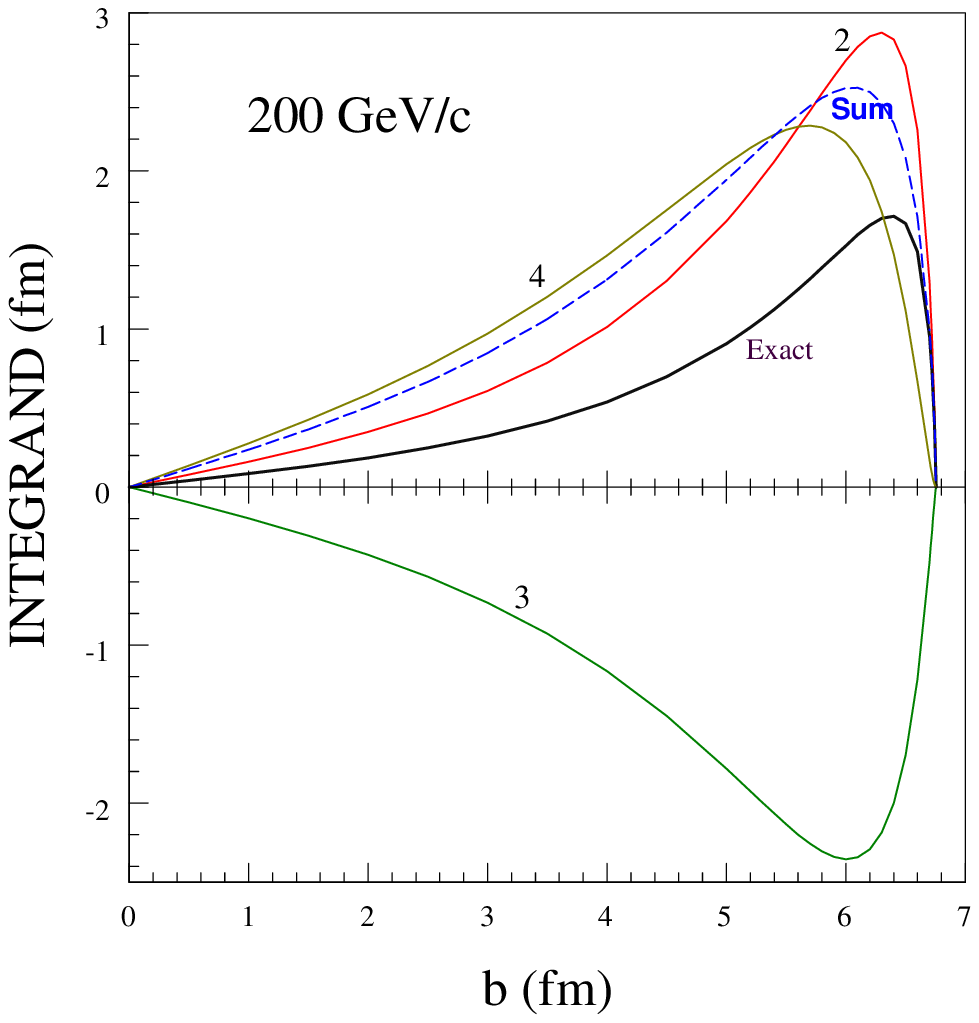}
    \caption{ Integrands for diffractive reductions in
    nucleon-$Pb^{208}$total cross sections as a function of impact parameter $b$
     at 200 GeV in 2nd through 4th order in the number of inelastic transitions.  The dashed curve
    shows the sum of these corrections, while the heavy solid curve shows the exact complete correction.}
\end{figure}

    The shapes of the integrands for some 50 and 200 GeV are shown in
Figs. 1 and 2, respectively, while the values of the diffractive
reductions in the total cross sections are given in Table 1. The
most surprising feature of these results is that, although the
contributions are all small compared to the un-corrected Glauber
cross section of about $2580 mb$ , at high energies the expansion
in the number of inelastic transitions does not converge well at
all.  As noted above, as the energy increases more and heavier
excited nucleons are included, and the correction at each order
increases. Because of the poor convergence, however, the expansion
in the number of inelastic transitions is not very useful,
especially at higher energies: it gives only an
order-of-magnitude for the exact result, with the individual
terms oscillating in sign.  The leading second order term in
particular is always of the right order of magnitude but larger
than the exact result, with the error increasing from 33\% at 50
GeV/c to nearly 80\% at 200 GeV/c.
\begin{table}
\caption{ Dispersive reductions in the total cross sections for
scattering of high-energy nucleons from $Pb^{208}$.  The figures
in parentheses are obtained if the longitudinal momentum transfers
are set to zero. }
 \vspace{.2cm}
 \begin{tabular} {||r||r|r|r|r|r||}  \hline
 Lab &  \multicolumn{5}{c||}{Dispersive Reductions in $
\sigma_{Total}(A=208) $ (mb)  ($\Delta p_{L} =0$)} \\
\cline{2-6} Mom &    \multicolumn{3}{c|}{Order in Inelastic
Transitions }&
\multicolumn{1}{c|}{}&\multicolumn{1}{c||}{} \\
 \cline{2-4}
 GeV  & \multicolumn{1}{c|}{2} & \multicolumn{1}{c|}{3}  &
\multicolumn{1}{c|}{4}
& \multicolumn{1}{c|}{Sum} & \multicolumn{1}{c||}{Exact}  \\
\hline\hline

 50  & 45.4 (57.9) & -16.4 (-39.2) & 5.6 (31.4)  & 34.6 (50.1) & 34.2 (37.1)
  \\ \hline
 100 & 59.6 (66.9) & -40.6 (-61.4) & 29.6 (65.7)  & 48.6 (71.2) &  37.6 (38.5) \\ \hline
 150 & 66.0 (70.6) & -56.9 (-73.0) & 54.4 (88.2)  & 63.5 (85.8)& 38.5 (38.9)\\ \hline
 200 & 69.7 (72.9) & -68.2 (-80.6) & 75.3 (104.8) & 76.8 (97.1) & 38.9 (39.2)\\ \hline

\end{tabular}

\end{table}

    The influence of the longitudinal momentum transfer was studied
by comparing the result calculated from the formulas above with
those with the longitudinal momentum transfers dropped (so that
$y_2$ becomes a real number).  The results  in this limit  are
given as the numbers in parenthesis in Table 1.  Although the
longitudinal momentum transfers reduce significantly the
magnitudes of the individual terms in the expansion, , they have
a relatively small effect,  decreasing with increasing energy, on
the exact results.

\section{Conclusion}

    Previous calculations of the dispersive corrections
considered here have considered only a small number of channels,
ignored the longitudinal momentum transfer, or included only the
lowest order term in the the transition expansion.  In the
context of our model it has been shown that all of these can lead
to large errors.
    It should be remembered, however, that the matrix model used above has many
 arbitrary features, even though it is roughly consistent with what is
 known about about high-mass single diffraction dissociation from a
 single nucleon.  In particular, the amplitudes for transitions from
 one highly excited nucleon state to another are essentially unknown, and the
 expressions used in the model are simply  guesses based  on the known
 behavior of the  amplitudes for excitation from the nucleon itself.  It
 would be useful to repeat the calculations with other assumptions for
 these amplitudes in order to get some idea of the dependence of the
 results on the assumptions.

    In the model used here the diffractive corrections to the
total cross sections are all small  compared to the total cross
section itself, but their expansion in the number of inelastic
transitions does not converge well at higher energies.  In
particular, the leading second order correction, which has been
used to estimate the size of the diffractive correction, is too
large by nearly 80\% at a laboratory momentum of 200 Gev/c.

    It would be interesting to extend these calculations to single
diffraction dissociation from nuclei, since for these processes
there
 is no zeroth order term, corresponding to the large Glauber contribution
 to the total cross section , so that the corrections due
to higher order terms might be relatively quite large.  A
preliminary investigation suggests that it should also be
possible to do an exact calculation in this case in the uniform
density limit.  (A calculation of single diffraction dissociation
from the deuteron would also be very interesting, and might put
additional constraints on the assumptions that go into the matrix
model.) ``Coherent'' diffraction dissociation, where the nucleus
remains in its ground state, would be particularly simple to
calculate, but probably experimentally challenging.   One could
also calculate a ``nuclear inclusive'' cross section in which all
nuclear excited states are summed over.

\section{Acknowledgement}

    The author would like to thank Professor Dieter Drechsel  and
    his colleagues in the Theory Group at the Institut f\"{u}r
    Kernphysik at the Univerit\"{a}t Mainz, where this work was
    completed, for their hospitality.

\bibliography{matmodlgRevTex4}  

\end{document}